\documentclass[aps,prl,superscriptaddress,twocolumn]{revtex4-1}

\usepackage{amsfonts}
\usepackage{amsmath}
\usepackage{multirow}
\usepackage{longtable}
\usepackage{graphicx}
\usepackage{color}
\usepackage{times}

\newcommand{\be}{\begin{equation}}
\newcommand{\ee}{\end{equation}}
\newcommand{\bea}{\begin{eqnarray}}
\newcommand{\eea}{\end{eqnarray}}
\newcommand{\av}[1]{\langle #1 \rangle}

\begin{document}

\title{The spreading of computer viruses on time-varying networks}

\author{Terry Brett}
\affiliation{University of Greenwich, Old Royal Naval College, London, UK}

\author{George Loukas}
\affiliation{University of Greenwich, Old Royal Naval College, London, UK}

\author{Yamir Moreno}
\affiliation{Institute for Biocomputation and Physics of Complex Systems (BIFI), University of Zaragoza, Zaragoza, Spain}
\affiliation{ISI Foundation, Turin, Italy}

\author{Nicola Perra}
\email[]{n.perra@greenwich.ac.uk}
\affiliation{University of Greenwich, Old Royal Naval College, London, UK}
\affiliation{ISI Foundation, Turin, Italy}
\date{\today}

\begin{abstract}
Social networks are the prime channel for the spreading of computer viruses. Yet the study of their propagation neglects the temporal nature of social interactions and the heterogeneity of users' susceptibility. Here, we introduce a theoretical framework that captures both properties. We study two realistic types of viruses propagating on temporal networks featuring Q categories of susceptibility and derive analytically the invasion threshold. We found that the temporal coupling of categories might increase the fragility of the system to cyber threats. Our results show that networks' dynamics and their interplay with users features are crucial for the spreading of computer viruses.
\end{abstract}

\maketitle

Alongside clear societal and economic benefits, modern technology exposes us to serious challenges. In particular, the spreading of malicious content online, often based on ingenious deception strategies, is one of the most pressing because it poses serious threats to our privacy, finances, and safety~\cite{kayes2017privacy}. Victims of a typical \emph{social engineering attack}~\cite{heartfield2016taxonomy} may receive a message containing a malicious link or file, appearing to originate from a friend or other trusted entity. If opened, it may compromise the computer, access personal information, and spread the virus further unbeknownst to the victim. Recent research has shown how the susceptibility of individuals to such attacks is not homogenous and depends on several features such as age, prior training, computer proficiency, familiarity with social network platforms, among others~\cite{heartfield2018detecting,heartfield2017eye,heartfield2016you}. Furthermore, the properties of real networks are known to facilitate the propagation of such processes~\cite{lloyd2001viruses,balthrop2004technological,satorras01-1,moreno02,newman02-4,newman10-1,pastor2015epidemic,barrat08-1,yang2013epidemics,yang2014spread}. In particular, the heterogeneity in contact patterns makes socio-technical systems quite fragile to biological and digital threats.

The study of these phenomena has largely neglected the complex temporal nature of online contact patterns in favor of static and time-aggregated approaches~\cite{holme2015modern,holme11-1}. These approximations might be fitting. Indeed, in the past, computer viruses would spread mainly via email networks, targeting the address books of victims, which contain contacts lists~\cite{newman2002email}. However, not many people create such lists any more and access to them is restricted~\cite{balthrop2004technological}. In the context of social or biological contagions, neglecting the temporal nature of the networks where the processes unfold has been shown to induce misrepresentations of their spreading potential. In fact, the order and concurrency of connections is key~\cite{barrat2015face,perra12-2,perra12-1,ribeiro12-2,PhysRevLett.112.118702,PhysRevE.87.032805,10.1063,starnini13-1,starnini_rw_temp_nets,valdano2015analytical,scholtes2014causality,Williams160196,rocha2014random,takaguchi2012importance,rocha2013bursts,ghoshal2006attractiveness,sun2015contrasting,mistry2015committed,pfitzner13-1,takaguchi12-1,takaguchi2013bursty,holme2014birth,holme2015basic,wang2016statistical,gonccalves2015social}.  To the best of our knowledge, beside some early work on the spreading of viruses via Bluetooth among mobile phones~\cite{wang2009understanding}, the study of the propagation of cyber threats considering the temporal nature of social interactions is still missing. Furthermore, with few exceptions~\cite{peng2017immunization}, the literature devoted to the study of computer viruses unfolding on networks typically neglects that the susceptibility of online users is not homogenous. Conversely, the literature that studies the susceptibility of users to cyber threats traditionally focuses on single users neglecting their connections. 

To tackle these limitations, here we introduce a theoretical framework to study the spreading of computer viruses, based on social engineering deception strategies, on time-varying networks. We model users' interactions using a time-varying network model and consider two types of viruses. The first mimics threats that can propagate only via connections activated at each time step. The second, on the contrary, considers viruses able to access also information about past connections. We investigate the impact of different classes of susceptibility considering that they might also influence the link formation process.  In all cases, we analytically derive the conditions regulating the spreading of the virus. Interestingly, these are defined by the interplay between the features of the cyber threats, the categories of susceptibility and their time-varying connectivity. Furthermore, in some scenarios, the coupling between categories creates a complex phenomenology that favors the spreading of the virus. These results have the potential to initiate future efforts aimed at describing more realistically the spreading of computer viruses on online social networks. 

We consider a population of $N$ online users which exchange messages in a time-varying network. Nodes are assigned to one of $Q$ categories describing their susceptibility to cyber threats measured in terms of their \emph{gullibility} and time needed to recover from successful attacks. Since susceptibility is linked to demographic features, we consider that the membership to a category might influence the link creation process. In fact, homophily is a strong social mechanism known to affect the structure and organization of ties~\cite{mcpherson2001birds}. We model the contact patterns between users with a generalization of the activity-driven framework~\cite{perra12-1,karsai13-1,ubaldi2015asymptotic,tizzani2018epidemic}. Here, nodes feature an activity $a$ describing their propensity to initiate communications. Activities are extracted from a distribution $F(a)$ which, as observations in real systems have shown, is typically heterogenous~\cite{perra12-1,ribeiro12-2,ubaldi2015asymptotic,tomasello2014role}. We select power-law distributions $F(a)\sim a^{-\alpha}$ with $a \in [\epsilon, 1] $ to avoid divergences. At each time step nodes are active with probability $a \Delta t$. Active nodes select $m$ others and create directed (out-going) links which mimic messages.  

In the simplest version of activity-driven networks the selection is random and memoryless~\cite{perra12-1}. Here, we propose a variation: with probability $p$ each target is selected, at random, among the group of nodes in the same category, and with probability $1-p$ among the nodes in any other category. In other words, $p$ tunes the homophily level in the network with respect to susceptibility to cyber threats. At time $t+\Delta t$ all edges are deleted and the process starts from the beginning. Unless specified otherwise, links have a duration $\Delta t$. Without loss of generality we set $\Delta t=1$. The model is clearly a simplification of real interactions. However, it offers simple, yet non trivial, settings to study the effects of temporal connectivity patterns on contagion processes unfolding at a comparable time-scale with respect to the evolution of connections~\cite{perra12-1,perra12-2,karsai13-1,liu2014controlling}.

We describe the propagation of a computer virus adopting the prototypical SIS model~\cite{keeling08-1,barrat08-1}. At each time step $t$ the virus, unbeknownst to the victims, sends a message, with malicious content, to all the nodes genuinely contacted at $t$ (virus type 1) or within $t-\tau$ time-step (virus type 2). The focus is not defining the optimal set of nodes to maximize/minimize the damage. Thus, we select randomly a small percentage ($0.5\%$) of nodes as initial seeds. In these settings, susceptible nodes of class $x \in [1,\ldots, Q]$, that receive a malicious message, become infectious with probability $\lambda_x$ which defines their gullibility. They recover and become susceptible again with rate $\mu_x$. Assuming that nodes with the same value of activity and in the same category are statistically equivalent, we group nodes according to the two features. At each time step, we call $S_a^x$ and $I_a^x$ the number of nodes susceptible and infected in activity class $a$ and category $x$. Clearly $\int da S_a^x=S^x$, $\int da I_a^x=I^x$, $\sum_x S^x=S$, and $\sum_x I^x=I$. Furthermore, $N_a^x$ describes the number of nodes of activity $a$ in category $x$, thus $\int da N_a^x=N^x$ and $\sum_{x}N^x=N$.  In these settings, we can represent the variation of the number of infected nodes of activity $a$ in category $x$ as:
\bea
\label{first}
&&d_t I_a^x= -\mu I_a^x+ \lambda_x m S_a^x \times \nonumber \\
&&\left [ p \int da' a' \frac{I_{a'}^x}{N^x} +(1-p)\sum_{y \neq x} \int da' a' \frac{I_{a'}^y}{N-N^y}  \right ].
 \eea
 
The first term on the right hand side accounts for the recovery process. The second and third terms capture susceptible nodes that receive messages from active and infected vertices in the same (second) or different (third) category, and get infected as a result. With respect to the typical biological contagion process, here transmission is asymmetric. Only nodes receiving a message from an infected person might be exposed to the virus. Thus, not only the order of connections, but also their direction is a crucial ingredient for the spreading. Since the links are created randomly, each node is selected with a probability $p m/N^x$ by nodes in the same category or $(1-p) m/(N-N^y)$ by nodes in other categories. The total number of nodes is constant thus $S_a^x=N_a^x-I_a^x$ and at the early stages of the spreading we can assume that the number of infected nodes is very small: $S_a^x\sim N_a^x$. By integrating across all activities Eq.~\ref{first} we get: 
\be
d_t I^x= -\mu_x I^x+ \lambda_x m \left [ p \theta^x +(1-p) N^x\sum_{y\neq x} \theta^y/(N-N^y)  \right ], \nonumber
\ee
where we define $\theta^x=\int da a I_a^x$. By multiplying both sides of Eq.~\ref{first} for $a$ and integrating across all the activities we obtain 
\bea
&&d_t \theta^x= -\mu_x \theta^x+ \nonumber\\
&&m \lambda_x \av{a}_x \left [ p \theta^x +(1-p) N^x\sum_{y\neq x} \theta^y/(N-N^y) \right ].\nonumber 
\eea
The virus is able to spread, if and only if the largest eigenvalue of the Jacobian matrix of the system of differential equations in $I^x$ and $\theta^x$ is larger than zero~\cite{perra12-1}. As shown in details in the Supplementary Material (SM) this implies:
\be
\label{general}
R_0=\frac{p\sum_{x}\beta_x+\Xi}{\sum_{x}\mu_x}>1,
\ee
where $R_0$ is the basic reproductive number defined as the average number of infected nodes generated, in a fully susceptible population, by an infected individual~\cite{keeling08-1}, $\beta_x=m\lambda_x \av{a}_x$ and $\Xi$ is a function of the interplay between the average activation, infection and recovery rate of each category as well as of the mixing between categories. 

To understand the dynamics, let us consider a particular case in which the system is characterized by only two categories. Furthermore, let us consider, as first scenario, that all nodes have the same recovery rate. In these settings we have
$\Xi^2=p^2(\beta_1+\beta_2)^2+4 \beta_1 \beta_2 (1-2p)$. The condition for the spreading, even with only two classes, is a non linear function of the average activity of each category, the infection probabilities per contact and the homophily. In the limit $p=0$, nodes in a category connects only with vertices in the other and the expression reduces to $R_0=\frac{\sqrt{\beta_1 \beta_2}}{\mu}$. In the limit $p=1$ instead, interactions are only between nodes in the same category. The system is effectively split in two disconnected networks and there are two independent conditions $R_0^x=\beta_x/\mu$. For a general $p$, these two values confine $R_0$: $\min_{x}R_0^x\le R_0(p)\le \max_{x}R_0^x$. In fact, any value of $p<1$ will reduce the spreading power of the category characterized with the largest $R_0^x$ as some connections will be established with nodes where the virus finds it harder to spread (see SM for the proof). 

\begin{figure}
\centering
\includegraphics[width=\columnwidth]{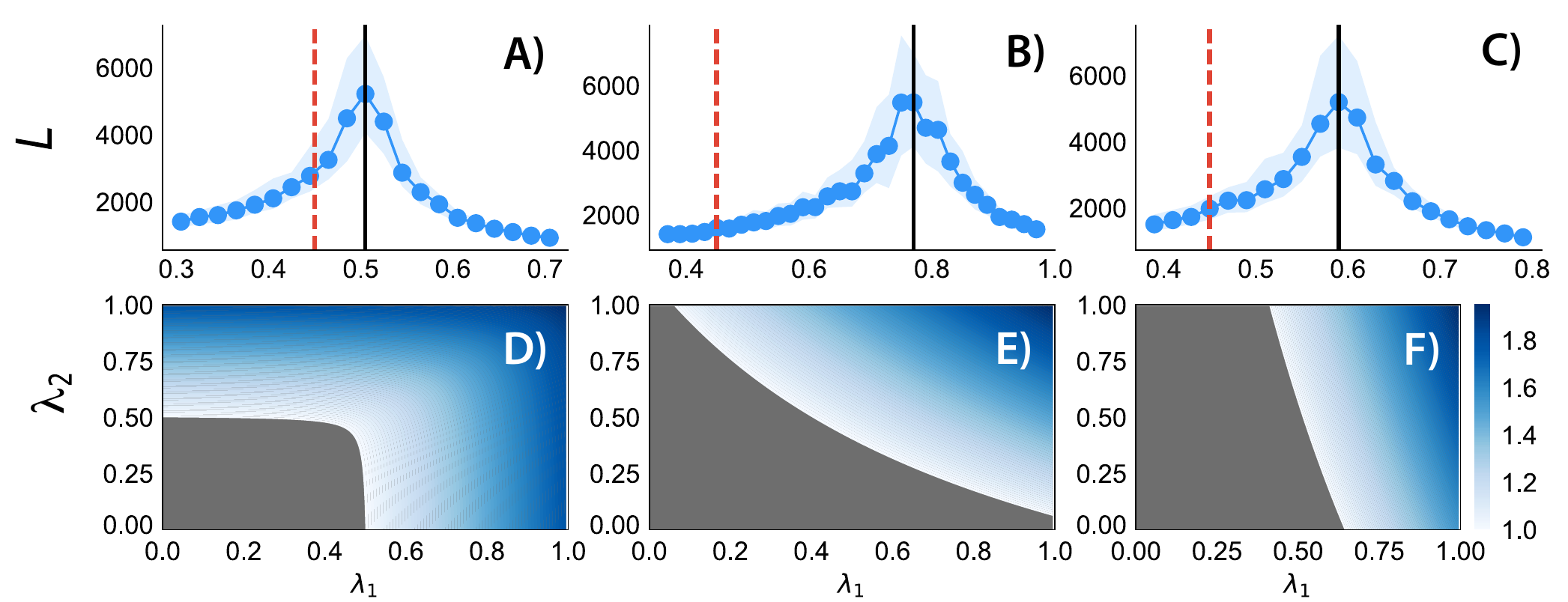}
\caption{Lifetime of the SIS process (A-C) and contour plot of $R_0$ (D-F). In A-B-D-E nodes are randomly assigned to two categories, in C-F instead in decreasing order of activity. We set $p=0.9$ (A-D), $p=0.4$ (B-C-E-F). In A-C we fix $N=2 \times 10^{5}$, $m=4$, $\alpha=-2.1$, $\mu_1=\mu_2=10^{-2}$, $\lambda_2=0.2$, $Y=0.3$, and $0.5\%$ of random initial seeds. We plot the median and $50\%$ confidence intervals in $10^2$ simulations per point. The solid lines come from Eq.~\ref{general}, and the dashed lines are the analytical threshold in case of a single category. In the contour plot we set $\mu_1=\mu_2=10^{-1}$.}
\label{fig:01}
\end{figure}

In Fig.~\ref{fig:01}-A-C, we compare analytical predictions with numerical simulations. We set $\lambda_2=0.2$ and use Eq.~\ref{general} to estimate the critical value of $\lambda_1$ for which $R_0\equiv1$. On the $y$-axis we plot the lifetime of the process defined as the time that the virus needs either to die out or to reach a fraction $Y$ of the population~\cite{boguna13-1}. The lifetime acts as the susceptibility of a second order phase transition and allows a precise numerical estimation of the threshold of SIS processes~\cite{boguna13-1}. In panels A-B we consider a scenario in which nodes are assigned randomly to one of the two categories. Thus the average activity in the two is the same and set $p=0.9$ and $p=0.4$ respectively. The analytical value of the threshold (vertical solid line) perfectly matches the numerical estimation. For $p=0.9$ the threshold is smaller than for $p=0.4$ and closer to the threshold of a system with a single category (dashed lines). For smaller values of homophily, instead, the critical conditions are driven by the interplay between the activation rates and gullibility of the two categories. Panels D-E show the analytical value of $R_0$ as a function of $\lambda_1$ and $\lambda_2$ for the two values of $p$. The grey regions are sub-critical, i.e., the virus is not able to spread. Since the average activity in the two categories is the same, the two plots are symmetric. Interestingly, the active region (where the virus is able to spread) is larger for large values of $p$. This is due to the fact that in these settings the virus will spread if above the threshold in at least one category independently of the other. In the opposite limit, on the contrary, the two categories get intertwined and a small value of the infection probability in one category should be associated to a progressively large value in the other. 

In panels C-F we consider that the first category contains a fraction $g$ of nodes selected in decreasing order of activity. Thus, this category contains the $gN$ most active nodes, while the other the $(1-g)N$ least active (see SM). To compare with panel B, we set  $g=0.5$ and $p=0.4$. First, the analytical threshold nicely matches the numerical simulations. Second, although the other parameters are the same used in panel B, the critical value of the gullibility of the first class is smaller. Thus, correlations between activity and gullibility facilitate the spreading. This is confirmed in panel F where the active phase space features a region in which the spreading is completely dominated by the category of most active nodes. Overall, all the plots show the importance of distinguishing nodes according to their gullibility. Indeed, neglecting the presence of different classes of users might induce a strong misrepresentation of the virus propagation (dashed lines).

\begin{figure}
\centering
\includegraphics[width=\columnwidth]{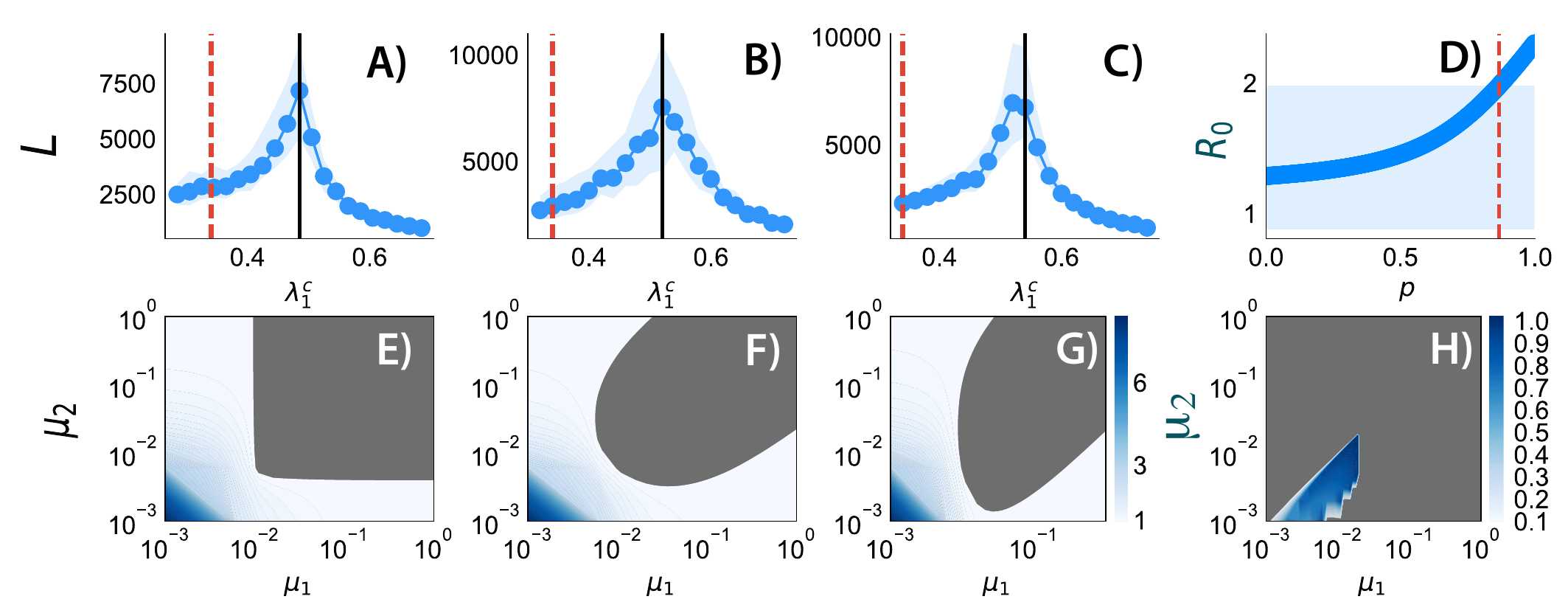}
\caption{Lifetime of the process (A-C), $R_0(\mu_1,\mu_2)$ (E-G), $R_0(p)$ (D), and $p^*(\mu_1,\mu_2)$ (H). In A-B-E-F nodes are randomly assigned to two categories, in C-G instead in decreasing order of activity. We set $p=0.9$ (A-E), $p=0.4$ (B-C-F-G). In panels A-C we set $N=2 \times 10^{5}$, $m=4$, $\alpha=-2.1$, $\mu_1=10^{-2}$, $\mu_2=5\times 10^{-3}$, $\lambda_2=0.2$, $Y=0.3$, and $0.5\%$ randomly selected seeds. We plot the median and $50\%$ confidence intervals in $10^2$ simulations per point. The solid lines come from Eq.~\ref{general}. The dashed lines are the analytical threshold in case of a single category of recovery rate characterized by the average value of the recovery rates. In the contour plot we set $\lambda_1=0.485$ and $\lambda_2=0.2$. In D the shaded area describe the region where $\min_{x}\beta_x/\mu_x \le R_0 \le \max_{x}\beta_x/\mu_x$. The dashed vertical line the the analytical value of $p$ above which $R_0>\max_{x}\beta_x/\mu_x$. In H we plot $p^*$ as function of $\mu_1$ and $\mu_2$. In D-H we set $\lambda_1=0.9$ and use the same parameters of the other plots.}
\label{fig:02}
\end{figure}

Let us next consider a second scenario where categories differentiate also for the time needed to recover from a successful attack. For two categories, we can write $\Xi^2=(\mu_1-\mu_2)^2+ p^2(\beta_1+\beta_2)^2+2p(\mu_2-\mu_1)(\beta_1-\beta_2)+4\beta_1 \beta_2 (1-2p)$. Interestingly, we have the same terms that appeared in the first scenario, plus two that feature the difference between the recovery rates and $\beta$s of the two categories. Thus $R_0$ is a function of the interplay between the activities, gullibilities and recovery rates. In the limit $p=0$, each category only connects with nodes in the other, the two groups are coupled and the threshold reads $R_0=\frac{\sqrt{(\mu_1-\mu_2)^2+4\beta_1 \beta_2}}{\mu_1+\mu_2}$. In the limit $p=1$ instead, the two categories are completely de-coupled and the threshold becomes, as before, $R_0=\beta_x/\mu_x$. As shown in Fig.~\ref{fig:02}-D-H, for a general value of $p$ the reproductive number is not bounded, as before, by the values of $R_0^x$ computed in the two classes separately (see SM). In Fig.~\ref{fig:02}-D, we assign nodes randomly to each category, fix $\beta_x$ and $\mu_x$ and compute $R_0$ as a function of $p$. In the shaded area $\min_x R_0^x \le R_0(p) \le \max_x R_0^x$. Interestingly, after a $p^*$ (vertical dashed line), which as shown in the SM can be computed analytically, we enter in a regime where $R_0 (p) > \max_x R_0^x$. Thus, only specific values of the coupling between categories might induce the virus to spread faster in the combined system than in each single category in isolation. However, this non linear effect is found only in a small fraction of the phase space see Fig.~\ref{fig:02}-H. The necessary, but not sufficient condition, is that two categories differentiate both for gullibility and recovery rates in such a way that one is more gullible and recovers faster than the other. In this regime, the right mixing between the two might create a feedback loop that makes the system more fragile.

Fig.~\ref{fig:02}-A-C shows a good match between the analytical (solid vertical lines) and numerical thresholds in case of nodes are assigned at random (A-B) or in decreasing order of activity (C) to the two categories. We fix two different recovery rates, $\lambda_2$, and use $\lambda_1$ as order parameter. Panels A-B differ in the value of the homophily $p$. We set $p=0.9$ in A, while $p=0.4$ in B-C. The presence of a category of nodes characterized by a smaller value of recovery rate pushes the threshold to smaller values with respect to the first scenarios (Fig.~\ref{fig:01}). As before, the value of the threshold estimated considering only a single category, characterized by the average recovery rate of the two, (dashed lines) leads to a misrepresentation of the spreading power of the virus, especially for smaller values of homophily (see panel B). 
 
The effect of $p$ on the critical value of $\lambda_1$ is similar to the first scenario. In fact, even when categories differentiate by the recovery rates, high values of homophily push the critical point to smaller values. However, here the difference between the two is less significant than in Fig.~\ref{fig:01}. In Fig.~\ref{fig:02}-E-F, we show the analytical value of $R_0$ as function of $\mu_1$ and $\mu_2$. Interestingly, the sub-critical region, for $p=0.4$, is smaller than for $p=0.9$. This is in contrast to what was observed in the corresponding plots for the first scenario and highlights once again the complex phenomenology introduced by the interplay of different recovery rates. In Fig.~\ref{fig:02}-C-G we investigate the effect of correlations. In case that the most active nodes are able to recover quickly from the attack, the virus is able to spread only if the gullibility of such users is higher than in the corresponding case without correlations (panel B). This is confirmed in panel G, where we see that correlations between recovery rates significantly change the active region.

\begin{figure}
\centering
\includegraphics[width=\columnwidth]{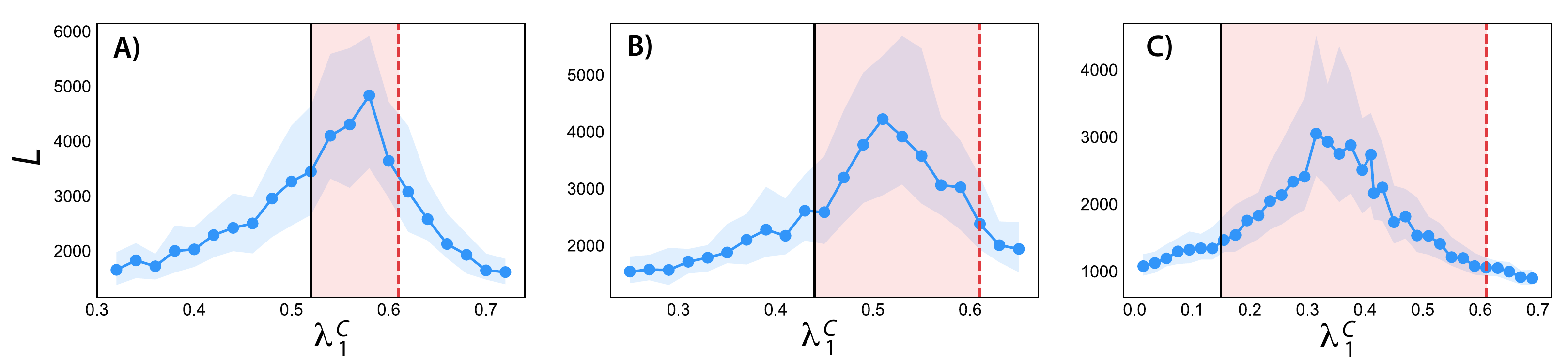}
\caption{Lifetime of the SIS process for $\tau=2,3,10$ (A,B,C) for two categories to which nodes are assigned randomly. Simulations are done setting $N=2 \times 10^{5}$, $m=4$, $\alpha=-2.1$, $Y=0.3$, $\mu=10^{-2}$, $\lambda_2=0.3$, $p=0.5$, and $0.5\%$ random initial seeds. We plot the median and $50\%$ confidence intervals in $10^2$ simulations per point. }
\label{fig:03}
\end{figure}

Finally, we turn our attention to a second type of virus able to access also past contacts of infected users within a time window $\tau$. As before, the virus propagates via active infected nodes, but at each time $t$ active users might infect their contacts in a time-window $(t-\tau,t]$. Within a mean-field approximation, we can adopt the same equations described above and change the probability that a node in each activity class receives a message by active and infected nodes. In this case, the out-degree of each active node is not $m$, but a function of $\tau$: $k^{out}(a)=m\left [ a+(\tau-1)a^2 \right ]$ (see SM). To grasp the derivation, consider the simplest scenario in which $\tau=2$. In this case, active nodes might have either $m$ or $2m$ contacts in two time steps. The first class describes nodes that are active at time $t$ but were not active at time $t-1$; whereas the second, nodes that were active in both time steps. Thus the out-degree of these nodes, on average, is $k^{out}(a)=ma(1-a)+2m a^2$. As shown in the SM, the condition for the spreading has the same structure of Eq.~\ref{general} where, however, the value of $\beta$s are changed with the following transformation $m \rightarrow m\left [ \av{a}+(\tau-1) \av{a^2} \right ]$. Thus, the larger the visibility of past connections, from the virus point of view, the larger $R_0$. Intuitively this is due to the fact that the virus, for large values of $\tau$, is able to access more contacts, which results in a larger spreading potential. This observation nicely shows how neglecting the temporal nature of connectivity patterns in favor of static (or time integrated) approximations might lead to a poor description of the propagation of viruses that do not have access to contacts lists or past connections. In Fig.~\ref{fig:03} we show the comparison between analytical (solid lines) and numerical values of the threshold for different values of $\tau$. To isolate the effect of $\tau$ we considered two categories, a single recovery rate, and set $p=0.5$. The analytical value is a good approximation only for small values of $\tau$. The mean-field approximation becomes less accurate as more connections from past time-steps are kept in memory. Thus, the analytical estimation provides only a lower bound, which together with the solution for $\tau=1$ (dashed lines) $-$that constitutes an upper bound$-$, marks the region where spreading is possible (red regions). In other words, for a general value of $\tau$, the threshold will be lower than the analytical value computed for $\tau=1$, and larger than the corresponding value computed at $\tau$.

Overall our results highlight how the spreading of computer viruses based on social engineering is critically affected by the temporal nature of our interactions and different susceptibilities to cyber threats. Our findings show that networks' dynamics and their interplay with the characteristics of users have to be considered in order to avoid misrepresentation of the spreading power of computer viruses in social networks. We have also quantified the extent to which the previous mismatch is important for three plausible scenarios. We, however, note that we have studied a simple network model that neglects a range of properties of real social networks such as the presence of weak and strong ties, high order correlations, and community structures. The study of the impact of these features on the unfolding of computer viruses calls for additional research.

This material is based upon work supported by, or in part by, the U. S. Army Research Laboratory and the U. S. Army Research Office under contract/grant number W911NF-18-1-0376. Y. M. acknowledges support from the Government of Arag\'on, Spain through grant E36-17R (FENOL) and by MINECO and FEDER funds (grant FIS2017-87519-P). The authors thanks Andrea Baronchelli and Michele Starnini for useful discussions. 

\bibliography{refs}

\section{Supplementary Material}

Here, we provide supplemental information about the mathematical derivation and present the sensitivity analysis of the results to the variation of the main parameters.

\section{Analytical derivations}

We consider two types of viruses. The first does not have access to the list of contacts of each victim, thus spread only via the connections activated at each time step $t$. The second instead, is able to access the list of contacts in the last $\tau$ time-steps. As we will see in details later, at the mean-field level, the structure of the equations regulating the variation of the number of infected individuals at early times is the same. However, for simplicity, let us consider first the first type of virus. \\
Each node is assigned to one of $Q$ categories $x \in [1,\ldots,Q]$ that distinguish nodes according the their gullibility $\lambda_x$, which describes the probability that they will not recognize the threat for example clicking on the added piece of malicious content, and the time they need to recover for a successful attack, $\mu_x^{-1}$ ($\mu_x$ is the recovery rate). Nodes are also characterized by their activity $a$ which describes their propensity to engage in social interactions in the unit time. To account for observations in real systems, we extract activity from a heterogenous distribution in particular a power-law $F(a)=B a^{-\alpha}$ with $a \in [\epsilon, 1] $ to avoid divergences. Each active node creates $m$ random connections. The target of each communication act is selected with probability $p$ within nodes in the same category and with probability $1-p$ with nodes in other categories. In both cases the actual target is selected at random. The virus will be able to spread only via the connections created by active and infected users. In particular, suppose that node $i$ has been compromised. At time $t$ the node activates and sends $m$ legitimate messages to $m$ users. During the same time-step, the virus, unbeknownst to $i$, will send a message with malicious content to all $m$ users. In these settings, the variation of number of infected nodes in each activity class $a$ and category $x$ can be written as:
\begin{widetext}
\be
\label{first}
d_t I_a^x= -\mu_x I_a^x+ \lambda_x m S_a^x\left [ p \int da' a' \frac{I_{a'}^x}{N^x} +(1-p)\sum_{y \neq x} \int da' a' \frac{I_{a'}^y}{N-N^y}  \right ].
 \ee
 \end{widetext}
The first term on the right hand side, describes the recovery process. The second term instead describes susceptible nodes that are connected by active nodes in the same category that are infected. These nodes get infected with probability $\lambda_x$ and selected with probability $p\frac{m}{N^x}$ (where $N^x$ is the total number of nodes in the category $x$). The third term, accounts for the same process but in which the susceptible node in activity class $a$ receives a message from active and infected nodes in other categories. Each node is selected with probability $(1-p)\frac{m}{N-N^y}$ by active vertices in class $y$. At early stages of the spreading we can assume that the number of infected to be very small respect to the susceptible thus we can approximate $S_a^x\sim N_a^x$. This is equivalent to neglect terms of the order of $(I_a^x)^2$. We can also define $\int da' a' I_{a'}^x=\Theta^x$, thus summing over all activity classes we get:
\be
d_t I^x= -\mu_x I^x+ \lambda_x m \left [ p \Theta^x +(1-p)\sum_{y \neq x} \frac{N^x}{N-N^y}\Theta^y  \right ].
 \ee
In order to characterize the behavior of the number of infected at such early times, we can write, starting from Eq.~\ref{first} the equation for each auxiliary function $\Theta^x$. In particular, we can multiply both sides of Eq.~\ref{first} for $a$ and integrate over all classes of activity. Doing so, we obtain:
\begin{widetext}
\be
d_t  \Theta^x= -\mu_x  \Theta^x+ \lambda_x m  \left [ p \Theta^x \int da \frac{a N_a^x}{N^x}+(1-p)\sum_{y \neq x} \frac{N^x}{N-N^y}\Theta^y \int da \frac{a N_a^x}{N^x}  \right ].
 \ee
 \end{widetext}
where we have multiply and divided the third term for $N^x$. We can now define $F_x(a)=\frac{N_a^x}{N^x}$ as the distribution of activities in the category $x$, and thus $\int da \frac{a N_a^x}{N^x}=\int da a F_x(a)=\av{a}_x$ is the average activity in the category. Finally, we let's define $c_{x,y}=\frac{N^x}{N-N^y}$ which is acts as the mixing probability between categorie. In these settings we get:
\be
d_t  \Theta^x= -\mu_x  \Theta^x+ \lambda_x m \av{a}_x \left [ p \Theta^x +(1-p)\sum_{y \neq x} c_{x,y} \Theta^y  \right ].
 \ee
 Thus we have a system of differential equations made of $2 Q$ equations. In particular, we have two equations for each $x$ in the form:
 \bea 
 d_t I^x &=& -\mu_x I^x+ \lambda_x m \left [ p \Theta^x +(1-p)\sum_{y \neq x} c_{x,y}\Theta^y  \right ] \nonumber \\ 
 &=& g^x. \nonumber \\
 d_t  \Theta^x &=& -\mu_x  \Theta^x+ \lambda_x m \av{a}_x \left [ p \Theta^x +(1-p)\sum_{y \neq x} c_{x,y} \Theta^y  \right ]\nonumber \\
  &=& h^x. 
 \eea
The conditions for the spreading can be identified by studying the eigenvalues of the Jacobian matrix of such system. The Jacobian can be written as follows:
\be
J=
\begin{bmatrix}
    \frac{\partial g^1 }{\partial I^1} & \frac{\partial g^1 }{\partial I^2} & \dots & \frac{\partial g^1 }{\partial I^Q}   & \frac{\partial g^1 }{\partial \Theta^1} & \frac{\partial g^1 }{\partial \Theta^2} & \dots & \frac{\partial g^1 }{\partial \Theta^Q} \\
    \frac{\partial g^2 }{\partial I^1} & \frac{\partial g^2 }{\partial I^2} & \dots & \frac{\partial g^2 }{\partial I^Q}   & \frac{\partial g^2 }{\partial \Theta^1} & \frac{\partial g^2 }{\partial \Theta^2} & \dots & \frac{\partial g^2 }{\partial \Theta^Q}\\
        \vdots & \vdots & \ddots & \vdots   &\vdots &\vdots & \ddots & \vdots \\
            \frac{\partial g^Q }{\partial I^1} & \frac{\partial g^Q }{\partial I^2} & \dots & \frac{\partial g^Q }{\partial I^Q}   & \frac{\partial g^Q }{\partial \Theta^1} & \frac{\partial g^Q }{\partial \Theta^2} & \dots & \frac{\partial g^Q }{\partial \Theta^Q}\\
                \frac{\partial h^1 }{\partial I^1} & \frac{\partial h^1 }{\partial I^2} & \dots & \frac{\partial h^1 }{\partial I^Q}   & \frac{\partial h^1 }{\partial \Theta^1} & \frac{\partial h^1 }{\partial \Theta^2} & \dots & \frac{\partial h^1 }{\partial \Theta^Q} \\
    \frac{\partial h^2 }{\partial I^1} & \frac{\partial h^2 }{\partial I^2} & \dots & \frac{\partial h^2 }{\partial I^Q}   & \frac{\partial h^2 }{\partial \Theta^1} & \frac{\partial h^2 }{\partial \Theta^2} & \dots & \frac{\partial h^2 }{\partial \Theta^Q}\\
        \vdots & \vdots & \ddots & \vdots   &\vdots &\vdots & \ddots & \vdots \\
            \frac{\partial h^Q }{\partial I^1} & \frac{\partial h^Q }{\partial I^2} & \dots & \frac{\partial h^Q }{\partial I^Q}   & \frac{\partial h^Q }{\partial \Theta^1} & \frac{\partial h^Q }{\partial \Theta^2} & \dots & \frac{\partial h^Q }{\partial \Theta^Q}\\
\end{bmatrix}
\ee
Substituting the general terms with the actual partial derivatives we get:
\begin{widetext}
\be
\label{Jacobian}
J=
\begin{bmatrix}
   -\mu_1 & 0 & \dots & 0 & p \lambda_1 m & (1-p) \lambda_1 m c_{1,2}  & \dots & (1-p) \lambda_1 m c_{1,Q} \\
   0 & -\mu_2 & \dots & 0 & (1-p) \lambda_2 m c_{2,1} & p \lambda_2 m  & \dots & (1-p) \lambda_2 m c_{2,Q} \\
        \vdots & \vdots & \ddots & \vdots   &\vdots &\vdots & \ddots & \vdots \\
   0 & 0 & \dots & -\mu_Q & (1-p) \lambda_2 m c_{Q,1} & (1-p) \lambda_2 m c_{Q,2}  & \dots & p \lambda_Q m \\
                0& 0 & \dots & 0   & -\mu_1 + p \beta_1& (1-p) \beta_1 c_{1,2} & \dots & (1-p) \beta_1 c_{1,Q} \\
     0& 0 & \dots & 0   & (1-p) \beta_2 c_{2,1} & -\mu_2 + p \beta_2 & \dots &  (1-p) \beta_2 c_{2,Q}\\
        \vdots & \vdots & \ddots & \vdots   &\vdots &\vdots & \ddots & \vdots \\
           0& 0 & \dots & 0   & (1-p) \beta_Q c_{Q,1} & (1-p) \beta_2 c_{Q,2} & \dots &  -\mu_Q + p \beta_Q\\
\end{bmatrix}
\ee
\end{widetext}
where we defined $\beta_x=m \av{a}_x \lambda_x$. It is important to notice the peculiarities of the Jacobian. The first $Q\times Q$ block made of the partial derivatives of the $g^x$ functions in the various $I^x$ is a diagonal block that features the recovery rates of each category. The second block on the bottom left side is a $Q \times Q$ block of all zeros. Indeed the variables $I^x$ do not appear in the $h^x$ equations. The adjacent block on the right, features in the diagonal the same function $-\mu_x + p\beta_x$. Due these properties, $Q$ eigenvalues are negative and equal to the negative of each recovery rate. The largest eigenvalue instead can be written as   
\be
\Lambda_{max}= -\sum_{x}\mu_x +p \sum_{x}\beta_x +\Xi
\ee
where $\Xi$ is an algebraic term function of all the $\beta_x$, $\mu_x$ and $c_{x,y}$. We focus on $\Lambda_{max}$ because the virus will be able to spread if and only if the largest eigenvalue is larger than zero. From this observation we obtain the conditions spreading:
\be
\label{general_r0}
R_0= \frac{p\sum_{x}\beta_x+ \Xi}{\sum_{x}\mu_x}>1
\ee
where $R_0$ is the reproductive number defined as the number of infected nodes generated by an initial seed in a fully susceptible population. It is important to mention that for any number of categories $\Xi$ has an analytical expression. However, since it derives from the characteristic equation of the Jacobian matrix, $\Xi$ gets more and more complicated as the dimensionality of the matrix increases. Generally speaking for $Q$ categories $\Xi$ is a polynomial of order $Q$ in all variables.
\subsection{Q=1}
In case of single category the expression of $R_0$ becomes:
\be
R_0=\frac{\beta}{\mu}
\ee
In fact, in this limit $p=1$ and the Jacobian matrix reduces to
\be
J=
\begin{bmatrix}
   -\mu & 0 \\
            0 & -\mu +\beta  \\
\end{bmatrix}
\ee
The two eigenvalues are $-\mu$ and $-\mu+\beta$. Thus the disease will be able to spread only if $\beta>\mu$.
\subsection{Q=2}
In the case of two categories, $Q=2$, the Jacobian becomes: 
\be
J=
\begin{bmatrix}
   -\mu_1 & 0 & p \lambda_1 m & (1-p) \lambda_1 m c_{1,2}\\
      0 & -\mu_2 & (1-p) \lambda_2 m c_{2,1} & p \lambda_2 m \\
            0 & 0 & -\mu_1 + p \beta_1 & (1-p) \beta_1 c_{1,2} \\
                    0 & 0 & (1-p) \beta_2 c_{2,1} & -\mu_2 + p \beta_2\\
\end{bmatrix}
\ee
In these settings we have:
\begin{widetext}
\be
\Xi^2=(\mu_1-\mu_2)^2+ p^2(\beta_1-\beta_2)^2 + 2p (\mu_2-\mu_1)(\beta_1-\beta_2)+ 4 \beta_1\beta_2 c_{1,2}c_{2,1}(p-1)^2
\ee
\end{widetext}
It is important to notice how with two categories, independently of their sizes $c_{1,2}=c_{2,1}=1$. In fact, the two sizes are constrained by $N=N^1+N^2$. Thus we have:
\begin{widetext}
\be
c_{1,2}=\frac{N^1}{N-N^2}=\frac{N^1}{N-N+N^1}=c_{2,1}=\frac{N^2}{N-N^1}=\frac{N^2}{N-N+N^2}=1
\ee
\end{widetext}
The expression of $\Xi$ reduces to: 
\begin{widetext}
\be
\Xi^2=(\mu_1-\mu_2)^2+ p^2(\beta_1+\beta_2)^2 + 2p (\mu_2-\mu_1)(\beta_1-\beta_2)+ 4 \beta_1\beta_2 (1-2p)
\ee
\end{widetext}
\subsection{$Q>2$}
As mentioned above, in the most general case of $Q$ categories, the expression of $\Xi$, becomes quite complex. However, its expression is set unequivocally by the characteristic equation of the Jacobian matrix and can be easily obtained with any programming language that allows symbolic computations such as Mathematica. The problem can be significantly simplified in case some of the variables describing the system are set. For example in the case of $Q=3$ one might wonder what is the critical value of $\lambda_1$ in a system in which $\beta_y$ (with $y=[2,3]$) and $\mu_y$ with ($y=[1,2,3]$) are set. In these settings, as shown later on, it is extremely easy to compute the largest eigenvalue of the Jacobian for the particular system under consideration as function of $\lambda_1$.

\subsection{$\tau>1$}

We now turn the attention to the second type of virus that is able to access not only the connections establish at time $t$ but also those in previous $\tau$ time steps. In order to characterize the conditions for the spreading in this case, let us first understand how many people del virus will be able to reach from each node of activity $a$. This number is equal to the out-degree of those nodes. In the case considered in the previous sections $\tau=1$, thus the virus was able to reach only the nodes contacted by each active and infected node within the time-step $t$. By construction, the out-degree of such nodes is $k^{out}(a)=ma$, since their are active with probability $a$ and when active they create $m$ random connections. What about for $\tau=2$?Active nodes at time $t$ might either have $m$ connections or $2m$. The first group describes nodes that were not active at time $t-1$ but they were active at time $t$. The second group instead describe nodes that were active in both time steps. Thus:
\be
k^{out}(a)= (1-a)am + 2m a^2= m(a + a^2).
\ee 
In fact, nodes of activity $a$ are not active with probability $1-a$ and are active two times in a row with probability $a^2$ (since the events are independent). The same reasoning applies for $\tau=3$. Here we could have three groups having either degree $m$, $2m$, and $3m$. As before, the first group describes nodes that were not active at time $t-2$ and $t-1$ but they were active at time $t$. The second group instead accounts for all the nodes that were active two times. Finally the third those that were active three times.  Thus we get:
\be
k^{out}(a)=m a (1-a)^2+ 4 m a^2 (1-a)+ 3m a^3= m (a+ 2 a^2)
\ee
In the case $\tau=4$ instead we have:
\bea
k^{out}(a) &=& m a (1-a)^3 + 6 m a^2 (1-a)^2 +\\ \nonumber
&+& 9 m a^3 (1-a)+ 4m a^4 \\ \nonumber
&=& m (a+ 3 a^2)
\eea
It is clear that the structure of the out-degree for a general $\tau$ can be written as: 
\be
k^{out}(a)=m\left [ a + (\tau-1) a^2\right ].
\ee
Within a mean-field approximation, we can approximate the process assuming that the virus will try to infected $k^{out}(a)$ other nodes as for the case $\tau=1$. This is an approximation because each active node, at time $t$, as a quenched list of contacts, those established in the time-steps before. The node will not re-draw them ex novo as in the case $\tau=1$. Thus, we can expect the approximation to be closer to the actual process for small values of $\tau$. Within such approach, the structure of the equation is the same as those above, the only different is in the $\beta$s since we will have $m \av{a}_x \rightarrow m\left [ \av{a}_x + (\tau-1) \av{a}_x^2\right ]$.
 
\section{Features of the phase space for $Q=2$}
Let's first consider the case in which $\mu_1=\mu_2=\mu$. The expression of $R_0$ reduces to:
\be
R_0=\frac{p(\beta_1+\beta_2)+\sqrt{p^2(\beta_1+\beta_2)^2+ 4\beta_1\beta_2 (1-2p)}}{2\mu}
\ee
In the limit $p=0$, nodes in each category will connect just with nodes in the other. The expression of $R_0$ becomes: $R_0=\sqrt{\beta_1\beta_2}/\mu$. In the opposite limit, $p=1$, nodes in the two categories are separated. Thus we have two independent conditions that have the same mathematical form we encountered for $Q=1$. In fact, we have $R_0^1=\beta_1/\mu$ and $R_0^2=\beta_2/\mu$. The virus will be able to spread in the system in case either of the $R_0^x$ are larger than one. Of course, in case both are larger than one each group will experience the virus. What happens in case $0<p<1$? It is interesting to notice how the value of $R_0$ for a general $p$ is bounded by the $R_0^x$ of the two categories taken in isolation: $\min_{x} R_0^x \le R_0(p) \le \max_x R_0^x$. Before the mathematical proof, let us try to develop the intuition behind. Suppose that $\beta_1>\beta_2$. Any value of $p<1$, will reduce the spreading power of nodes in the first category. In fact, nodes in category one will be connected to some nodes in category two that are less gullible, or less active, or create a smaller number of connection (remember that $\beta_x=m\av{a}_x\lambda_x$). Conversely, nodes in category two, will get in contact with nodes that increase the spreading potential of the virus. In order to prove this, let us consider the case $\beta_1>\beta_2$. We have to show how $R_0^1> R_0(p)$ and $R_0^2 < R_0(p)$. Let us consider the first condition: 
\be
\label{cond1}
\frac{\beta_1}{\mu}> \frac{p(\beta_1+\beta_2)+\sqrt{p^2(\beta_1+\beta_2)^2+ 4\beta_1\beta_2 (1-2p)}}{2\mu},
\ee
which is equivalent to:
\be
\beta_1(2-p)- p \beta_2 > \sqrt{p^2(\beta_1+\beta_2)^2+ 4\beta_1\beta_2 (1-2p)}
\ee
This condition is respected in case $\beta_1(2-p)- p \beta_2 > 0$, $p^2(\beta_1+\beta_2)^2+ 4\beta_1\beta_2 (1-2p) > 0$ and $(\beta_1(2-p)- p \beta_2)^2 > p^2(\beta_1+\beta_2)^2+ 4\beta_1\beta_2 (1-2p)$. The first condition implies $\beta_1 > \frac{p\beta_2}{2-p}$, which is always true since $\beta_1>\beta_2$ was the initial assumption. Furthermore, it is easy to show that equation $p^2(\beta_1+\beta_2)^2+ 4\beta_1\beta_2 (1-2p)= 0$ as no solution in $p$, thus the condition is always respected. Finally, the third condition implies
\begin{widetext}
\be
4\beta_1^2+ p^2 \beta_1^2 - 4\beta_1^2 p +p^2\beta_2^2 -2 p (2-p)\beta_1\beta_2 > p^2\beta_1^2 + p^2\beta_2 + 2p^2\beta_1\beta_2+ 4\beta_1\beta_2 - 8p\beta_1\beta_2
\ee
\end{widetext}
that reduces to $\beta_1 > \beta_2$. The three conditions prove Eq.~\ref{cond1} for all $p$. We have now to prove
\be
\label{cond2}
\frac{\beta_2}{\mu} < \frac{p(\beta_1+\beta_2)+\sqrt{p^2(\beta_1+\beta_2)^2+ 4\beta_1\beta_2 (1-2p)}}{2\mu},
\ee
which is equivalent to:
\be
\beta_2(2-p)- p \beta_1 < \sqrt{p^2(\beta_1+\beta_2)^2+ 4\beta_1\beta_2 (1-2p)}
\ee
This condition is respected in region in which $\beta_2(2-p)- p \beta_1 \ge 0$, $(\beta_2(2-p)- p \beta_1)^2 < p^2(\beta_1+\beta_2)^2+ 4\beta_1\beta_2 (1-2p)$ and $p^2(\beta_1+\beta_2)^2+ 4\beta_1\beta_2 (1-2p) \ge 0$, $\beta_2(2-p)- p \beta_1 < 0$. The first two conditions are respected when in the region $\frac{p\beta_1}{2-p}\le \beta_2 < \beta_1$. The other two instead in the region $\beta_2 < \frac{p\beta_1}{2-p}$. Overall, Eq.~\ref{cond2} is valid in the union of these two that implies $\beta_2 < \beta_1$ which is exactly the initial assumption. \\

\begin{figure}
\centering
\includegraphics[scale=0.28]{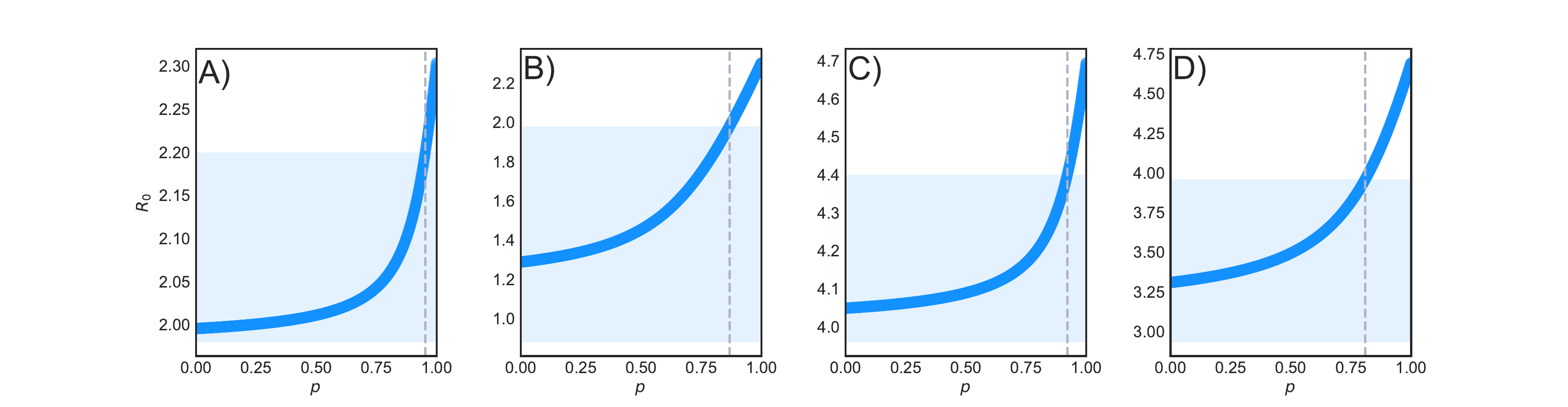}
\caption{ $R_0$ as function of $p$. The shaded area describe the region in which $\min_x \beta_x/\mu_x \le R_0 \le \max_x \beta_x/\mu_x$. The vertical line describe the value of $p^*$ from conditions Eq.~\ref{cond_mu2} and Eq.~\ref{cond_mu1}. In panels A-B we set $\mu_1=10^{-2}$, $\mu_2= 5 \times 10^{-3}$, $m=4$, $\lambda_1=0.9$,$\lambda_2=0.5$ (A) and $\lambda_2=0.2$ (B). In panels C-D we set $\mu_1=5 \times 10^{-3}$, $\mu_2= 3 \times 10^{-3}$, $m=4$, $\lambda_1=0.9$,$\lambda_2=0.6$ (C) and $\lambda_2=0.4$ (D).  }
\label{SI_fig:01}
\end{figure}

Let's consider now the general case in which also the two recovery rates are different. In the limit $p=0$, we have $R_0=\frac{\sqrt{(\mu_1-\mu_2)^2-4\beta_1\beta_2}}{\mu_1+\mu_2}$. In the opposite limit instead, $p=1$, the two categories are independent thus we have two conditions as before: $R_0^1=\beta_1/\mu_1$ and $R_0^2=\beta_2/\mu_2$. It is interesting to notice how in case the two recovery rates are not the same, the phase space of the process becomes significantly more complex. In fact, differences in the rate at which nodes recovers might create interesting non-linear behaviors. In particular, consider a scenario in which the first category features a larger $\beta_1$ and $\mu_1$ respect to the second. Thus, such nodes are more prone to infection but recover faster. In case $p<1$, the coupling between the two categories might boost the spreading of the virus, since the node in category one are able to infect those in two which, although less prone to the disease stay infected for longer. For a given configuration of parameters (i.e. setting $\beta$s and $\mu$s) we can analytically determine the value of $p$ above which this phenomenon is observed. In particular, let's assume that $\beta_1/\mu_1< \beta_2/\mu_2$. Next, we need to compute the value of $p$ (if any), for which $\beta_2/\mu_2< R_0(p)$. This implies:
\be
\frac{\beta_2}{\mu_2}< \frac{p(\beta_1+\beta_2)+\Xi}{\mu_1+\mu_2}
\ee
that can be written as:
\begin{widetext}
\be
\beta_2(\mu_1+\mu_2)- \mu_2 p(\beta_1+\beta_2)< \mu_2 \sqrt{(\mu_1-\mu_2)^2+ p^2(\beta_1+\beta_2)^2+2p(\mu_2+\mu_1)(\beta_1-\beta_2)+4 \beta_1 \beta_2 (1-2p)}
\ee
\end{widetext}
It is important to notice how this inequality is at the first order in $p$. Indeed, all second order terms cancel out. The value of $p$ that verifies the above inequality lays in the union of two systems of inequalities: i)$\beta_2(\mu_1+\mu_2)- \mu_2 p(\beta_1+\beta_2)<0$ and the quantity inside the square root is larger equal than zero, ii) $\beta_2(\mu_1+\mu_2)- \mu_2 p(\beta_1+\beta_2)>0$, and $(\beta_2(\mu_1+\mu_2)- \mu_2 p(\beta_1+\beta_2))^2<\mu_2^2\Xi^2$. Extensive numerical computations show that the values inside the square roots are always positive. Furthermore, the first condition in the first system result in values of $p$ always larger than one. Thus, the first system does not provide any physical ($p<1$) condition. Conversely, the first condition in the second system implies $p<1$ while the second:
\begin{widetext}
\be
\label{cond_mu2}
p>p^*=\frac{\beta_2^2(\mu_2+\mu_1)^2-\mu_2^2(\mu_1-\mu_2)^2-4\beta_1\beta_2 \mu_2^2}{2\mu_2\beta_2 (\mu_2+\mu_1)(\beta_1+\beta_2)+2\mu_2^2(\mu_2-\mu_1)(\beta_1-\beta_2)-8\beta_1\beta_2\mu_2^2}.
\ee
\end{widetext}
Thus, this is the only physical condition necessary to observe a reproductive number larger than in each category in isolation. Clearly, in the case $\beta_2/\mu_2<\beta_1/\mu_1$ the condition above becomes:
\begin{widetext}
\be
\label{cond_mu1}
p>p^*=\frac{\beta_1^2(\mu_2+\mu_1)^2-\mu_1^2(\mu_1-\mu_2)^2-4\beta_1\beta_2 \mu_1^2}{2\mu_1\beta_1 (\mu_2+\mu_1)(\beta_1+\beta_2)+2\mu_1^2(\mu_2-\mu_1)(\beta_1-\beta_2)-8\beta_1\beta_2\mu_1^2}.
\ee
\end{widetext}
In Figure~\ref{SI_fig:01} we verify the above condition. In particular, we set the values of $\beta_x$ and $\mu_x$ and plot $R_0$ from Eq.~\ref{general_r0} as function of $p$. In particular, we consider that nodes are assigned to the categories randomly. The shaded area is the region where $\min_x \beta_x/\mu_x \le R_0 \le \max_x \beta_x/\mu_x$. The vertical line show the value of $p^*$ determined from the condition derived above. It is clear how for a given setting, there might be a value of $p$ above which the reproductive number gets indeed larger than the the $R_0^x$ of each category in isolation. 

\begin{figure}
\centering
\includegraphics[scale=0.3]{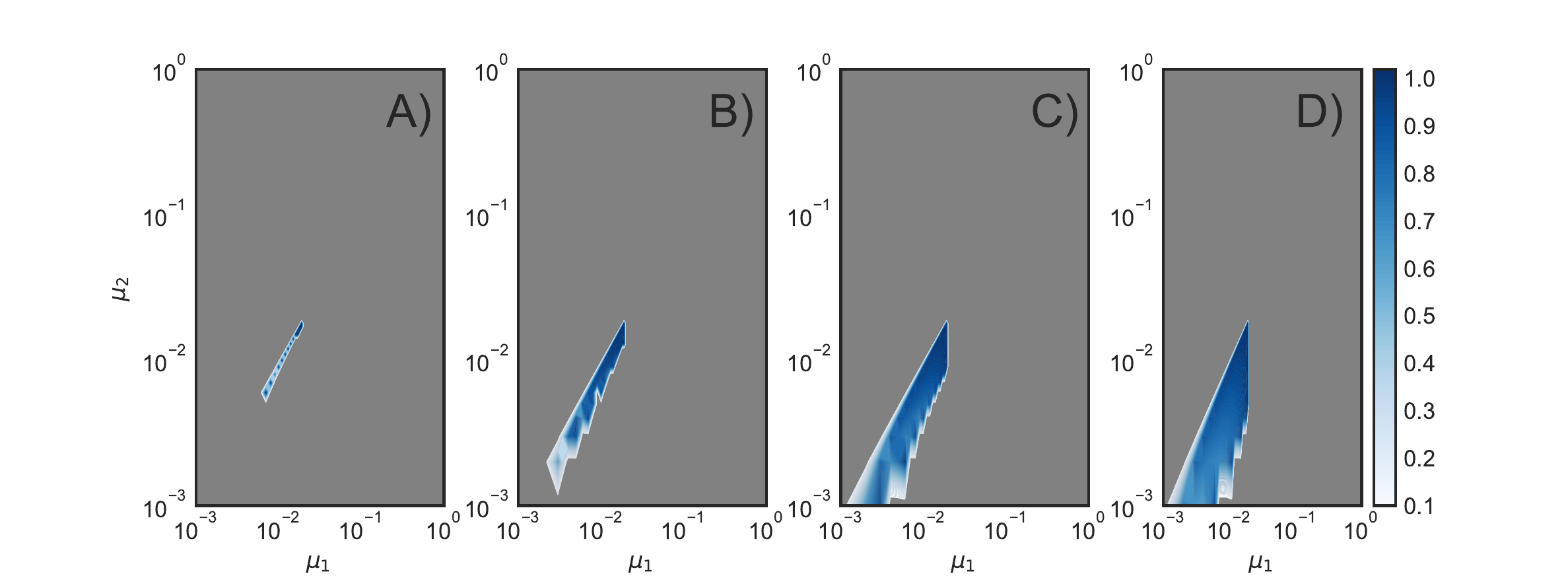}
\caption{We show as function of $\mu_1$ and $\mu_2$ the region of parameters in which the reproductive number of system is larger than the correspondent values computed in each category in isolation. The colors refer to the value of $p$ (calculated from Eq.~\ref{cond_mu2} and Eq.~\ref{cond_mu1}) above which this phenomenon is observed. We set $\lambda_1=0.9$, $\lambda_2=0.8$ (A),  $\lambda_2=0.6$ (B),  $\lambda_2=0.4$ (C),  $\lambda_2=0.2$ (D)  }
\label{SI_fig:02}
\end{figure}

It is important to stress how the region of the phase space in which we observe this phenomenon is generally speaking quite limited. In fact, it might happen only in case the category with the larger recovery rates has also the larger gullibility. In Figure~\ref{SI_fig:02} we show as contour plots the region of parameters where the reproductive number of the system is larger than that correspondent value in the two categories in isolation. In particular, we set $\lambda_1=0.9$, $\lambda_2=0.8$ (A),  $\lambda_2=0.6$ (B),  $\lambda_2=0.4$ (C),  $\lambda_2=0.2$ (D) and show as function of $\mu_1$ and $\mu_2$ the value of $p^*$. It is clear this region increases as the difference between the two gullibilities increases.
It is important to notice how the expression for $p^*$ is perfectly in line with the case in which $\mu_1=\mu_2$. Indeed, in this limit we get $p^*>1$ which implies, as expected, that the necessary condition to have a reproductive number larger than in each category in isolation is to have different recovery rates. 

\section{Numerical simulations}

\begin{figure}
\centering
\includegraphics[scale=0.35]{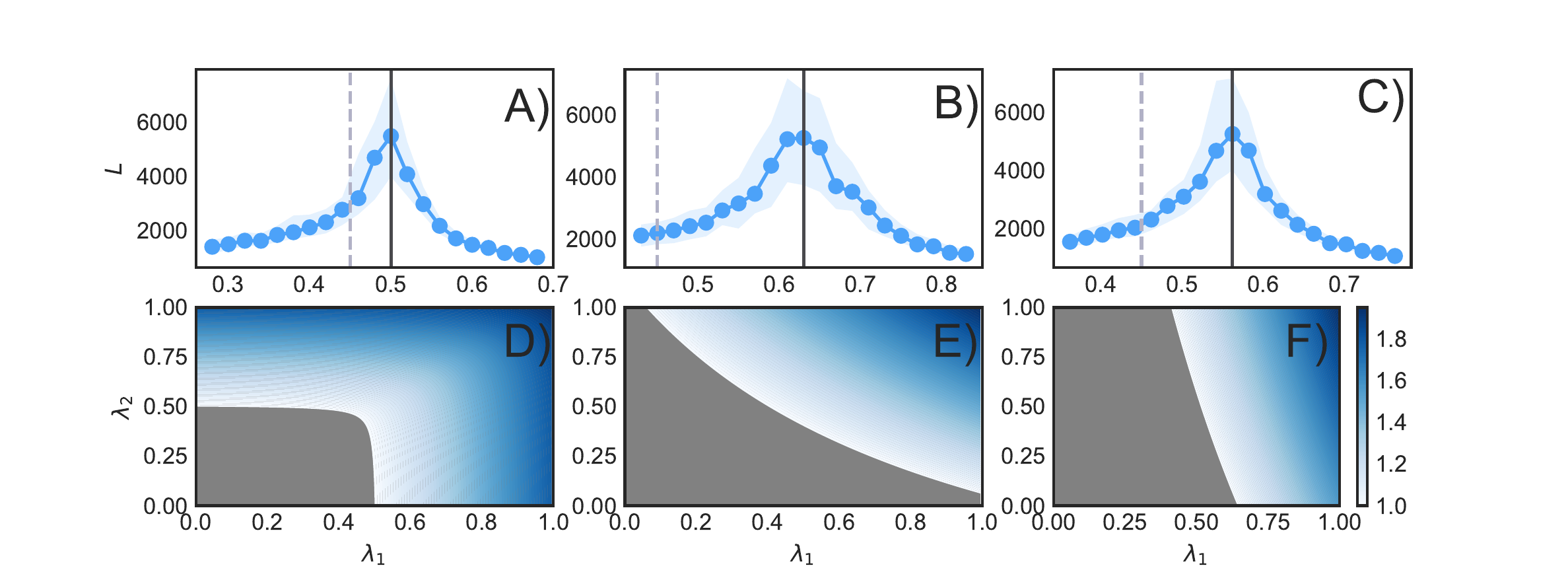}
\caption{Lifetime of the SIS process (A-C) and contour plot of $R_0$ (D-F). In A-B-D-E nodes are randomly assigned to two categories, in C-F instead in decreasing order of activity. We set $p=0.9$ (A-D), $p=0.4$ (B-C-E-F). In A-C we fix $N=2 \times 10^{5}$, $m=4$, $\alpha=-2.1$, $\mu_1=\mu_2=10^{-2}$, $\lambda_2=0.3$, $Y=0.3$, and $0.5\%$ of random initial seeds. We plot the median and $50\%$ confidence intervals in $10^2$ simulations per point. The solid lines come from Eq.~\ref{general_r0}, and the dashed lines are the analytical threshold in case of a single category. In the contour plot we set $\mu_1=\mu_2=10^{-1}$.}
\label{SI_fig:03}
\end{figure}

\begin{figure}
\centering
\includegraphics[scale=0.35]{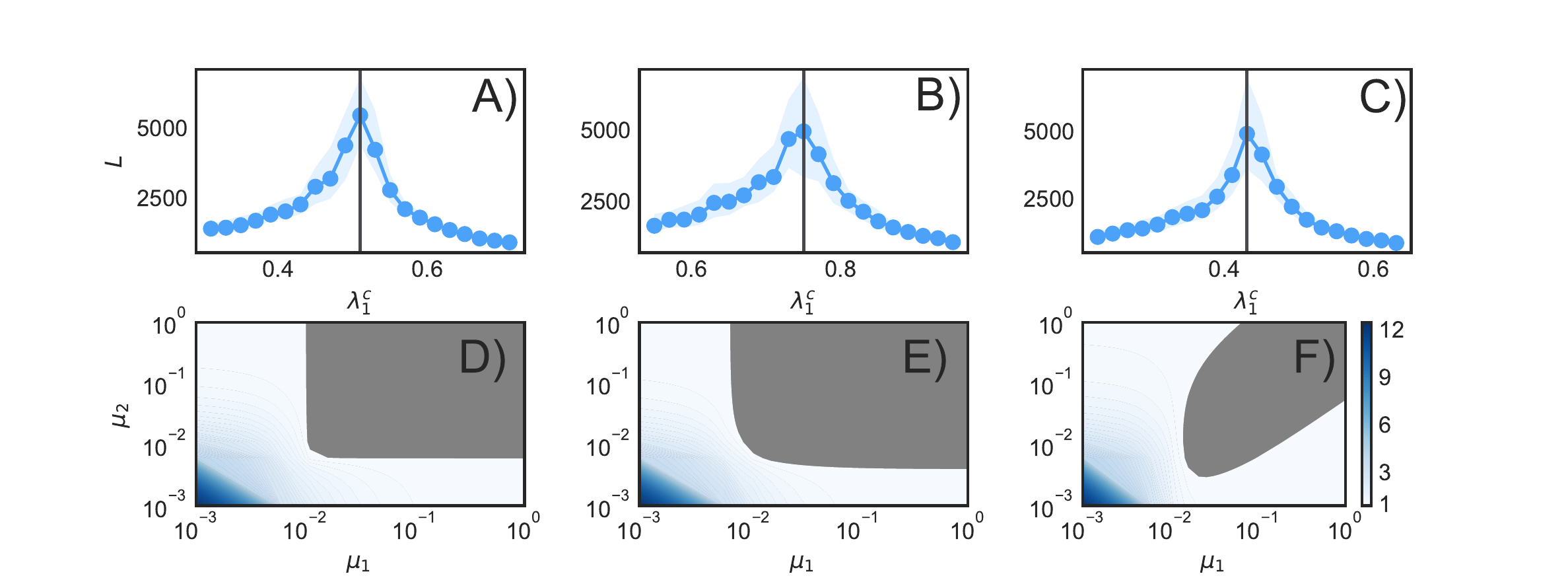}
\caption{Lifetime of the SIS process (A-C) and contour plot of $R_0$ (D-F). In A-B-D-E nodes are randomly assigned to two categories, in C-F instead in decreasing order of activity. We set $p=0.9$ (A-D), $p=0.6$ (B-C-E-F). In A-C we fix $N=2 \times 10^{5}$, $m=4$, $\alpha=-2.1$, $\mu_1=10^{-2}$, $\mu_2=10^{-1}$, $\lambda_2=0.3$, $Y=0.3$, and $0.5\%$ of random initial seeds. We plot the median and $50\%$ confidence intervals in $10^2$ simulations per point. The solid lines come from Eq.~\ref{general_r0}, and the dashed lines are the analytical threshold in case of a single category. In the contour plot we set $\lambda_1=0.51$ and $\lambda_2=0.3$.}
\label{SI_fig:04}
\end{figure}

\begin{figure}
\centering
\includegraphics[scale=0.35]{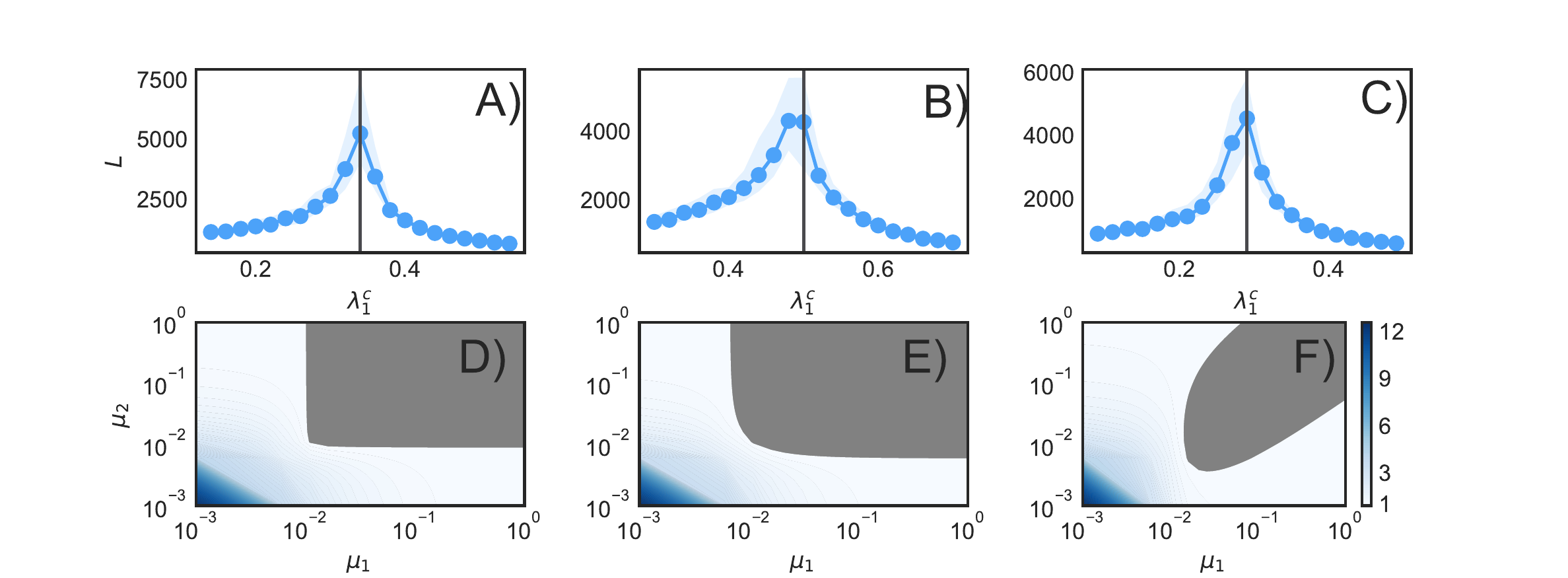}
\caption{Lifetime of the SIS process (A-C) and contour plot of $R_0$ (D-F). In A-B-D-E nodes are randomly assigned to two categories, in C-F instead in decreasing order of activity. We set $p=0.9$ (A-D), $p=0.6$ (B-C-E-F). In A-C we fix $N=2 \times 10^{5}$, $m=6$, $\alpha=-2.1$, $\mu_1=10^{-2}$, $\mu_2=10^{-1}$, $\lambda_2=0.3$, $Y=0.3$, and $0.5\%$ of random initial seeds. We plot the median and $50\%$ confidence intervals in $10^2$ simulations per point. The solid lines come from Eq.~\ref{general_r0}, and the dashed lines are the analytical threshold in case of a single category. In the contour plot we set $\lambda_1=0.34$ and $\lambda_2=0.3$.}
\label{SI_fig:05}
\end{figure}

\begin{figure}
\centering
\includegraphics[scale=0.35]{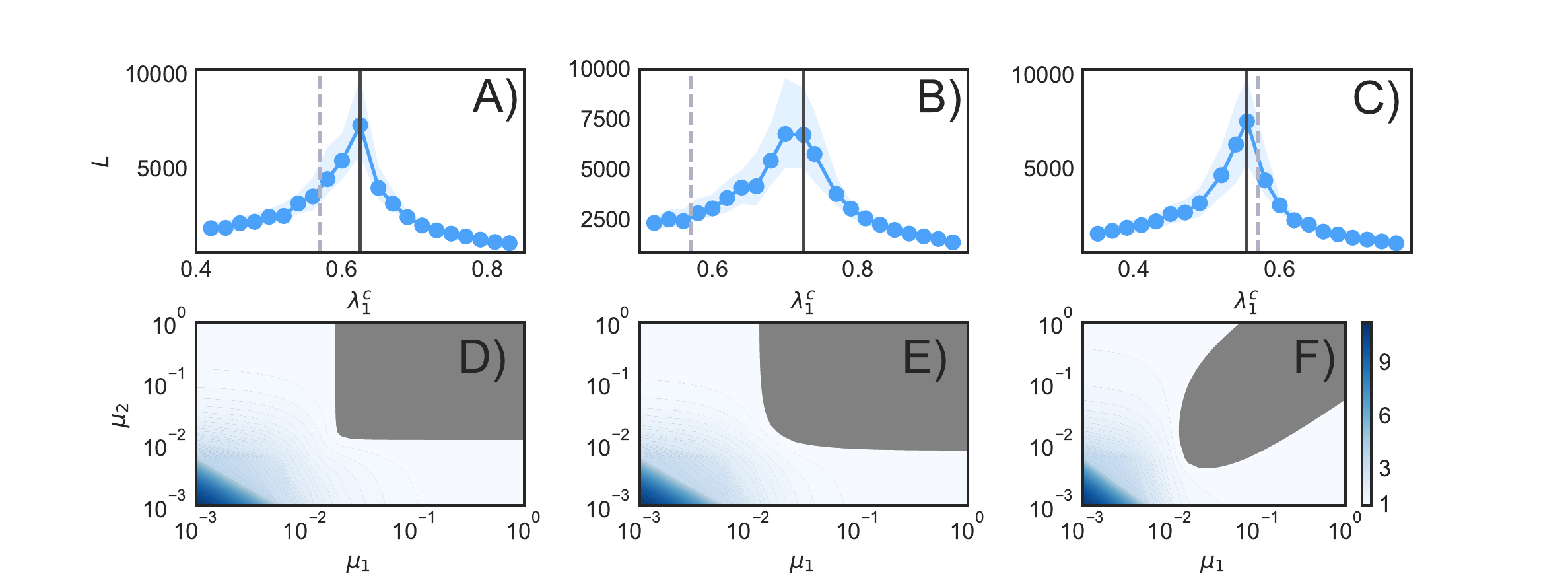}
\caption{Lifetime of the SIS process (A-C) and contour plot of $R_0$ (D-F). In A-B-D-E nodes are randomly assigned to two categories, in C-F instead in decreasing order of activity. We set $p=0.9$ (A-D), $p=0.6$ (B-C-E-F). In A-C we fix $N=2 \times 10^{5}$, $m=6$, $\alpha=-2.5$, $\mu_1=10^{-2}$, $\mu_2=10^{-2}$, $\lambda_2=0.4$, $Y=0.3$, and $0.5\%$ of random initial seeds. We plot the median and $50\%$ confidence intervals in $10^2$ simulations per point. The solid lines come from Eq.~\ref{general_r0}, and the dashed lines are the analytical threshold in case of a single category. In the contour plot we set $\lambda_1=0.625$ and $\lambda_2=0.5$.}
\label{SI_fig:06}
\end{figure}

In this section we will present the sensitivity analysis to the model's parameters. We will first consider two categories ($Q=2$). As shown in the main text, we adopted two main approaches to assign node to categories. The first is at random, the second is instead in decreasing order to activity. In particular, we order activity in decreasing order and then assign the first $gN$ nodes to the first category and the remaining to the second. Thus $\av{a}_1=\int_{a_c}^{1} da a F(a)$ and $\av{a}_2=\int_{\epsilon}^{a_c} da a F(a)$ and $a_c$ is determined in such a way that the fraction of nodes in the first class is $g$. This can be easily done imposing:
\be
\int_{a_c}^1 F(a)da = g.
\ee
Since $F(a)=\frac{1-\alpha}{1-\epsilon^{1-\alpha}} a^{-\alpha}$ we get: 
\be
a_c=\left [1-g (1-\epsilon^{1-\alpha}) \right ]^{\frac{1}{1-\alpha}}
\ee
It is important to notice that in Eq.~\ref{general_r0} the expression of $\av{a}_x$ in the two assignment scenarios is slightly different. In particular we defined $\av{a}_x= \int da F_x(a) a =\int da \frac{N_a^x}{N^x}a$. In case nodes are assigned randomly to the two categories we have that $F_x(a)\sim F(a)$ since $N_a^x=N_a/g$ and $N^x=N/g$ (where $g$ is the fraction of node in the general category $x$ in this case). Thus, $\av{a}_x=\av{a}$ for the two categories. In case instead nodes are assigned in decreasing order of activity $\av{a}_x=\av{a}/g$. In fact, in this limit $N_a^x=N_a$ (since nodes are assigned to categories as function of their activity) but $N^x=gN$. \\
In Figure~\ref{SI_fig:03}, we consider the case of $\mu_1=\mu_2$ in case of randomly selected nodes for $p=0.9$ (A-D) and $p=0.4$ (B-E). We also consider the case of correlation between category and activity for $p=0.4$ (C-F). Respect to Figure 1 of the main text, we used a different value of $\lambda_2=0.3$. Across the board the analytical solution match perfectly the numerical estimation of the threshold. \\
In Figure~\ref{SI_fig:04} we consider the general case of different recovery rates. In particular we consider a different set of values respect to those in the main text. In particular, we set $\mu_1=10^{-2}$, $\mu_2=10^{-1}$, $\lambda_2=0.3$ and $m=4$. In panels A-B-D-E we consider random assignment of nodes to categories. In C-F we consider the correlation between activity and category. We assign to category one to most active nodes. Also, in panels B-C-E-F we considered $p=0.6$ while in panel A-D we set $p=0.9$. Overall, the figure confirms the validity of the theoretical approach and highlights one more time the effects of correlations between category assignment and activity that reduce the non-active phase space (see panel F). Furthermore, it is important to notice how the critical value in case of a single category with a recovery rate average of the two here would be $\lambda_1^c=2.5$ (not shown in the figure) which implies that the virus would not be able to spread since all the gullibilities should be smaller or equal to $1$. This confirms the importance of accounting for the presence of different categories of users in order to correctly capture the spreading power of the virus.  \\
In Figure~\ref{SI_fig:05} we test the sensitivity to the parameter $m$. In the main text as well as in many of the other plots we set $m=4$. Here, we fix instead $m=6$ keeping all the other parameters the same as in the Figure~\ref{SI_fig:04}. The analytical solutions one more time match the numerical simulations and the contour plots confirm the picture discussed in the main text and all the other similar plots.
In Figure~\ref{SI_fig:06} we test the sensitivity to the exponent of the activity distribution. In all the other plots we set $\alpha=-2.1$, here instead we consider $\alpha=-2.5$. We considered a scenario in which the recovery rates of the two categories is the same, set $\lambda_2=0.4$, $m=6$ and consider two different values of $p$. As clear from the figure, also in this case the analytical  estimation matches the numerical simulations. Furthermore, it is interesting to notice how, in case of faster decay of the activity distribution (i.e. smaller value of the exponent $\alpha$), the threshold of the correlated case (panel C-F) is closer to the scenario of a single category (dashed line). Indeed, the average activity of the more active category gets closer to the average activity of the whole network.   

\subsection{$Q=3$}
\begin{figure}
\centering
\includegraphics[scale=0.35]{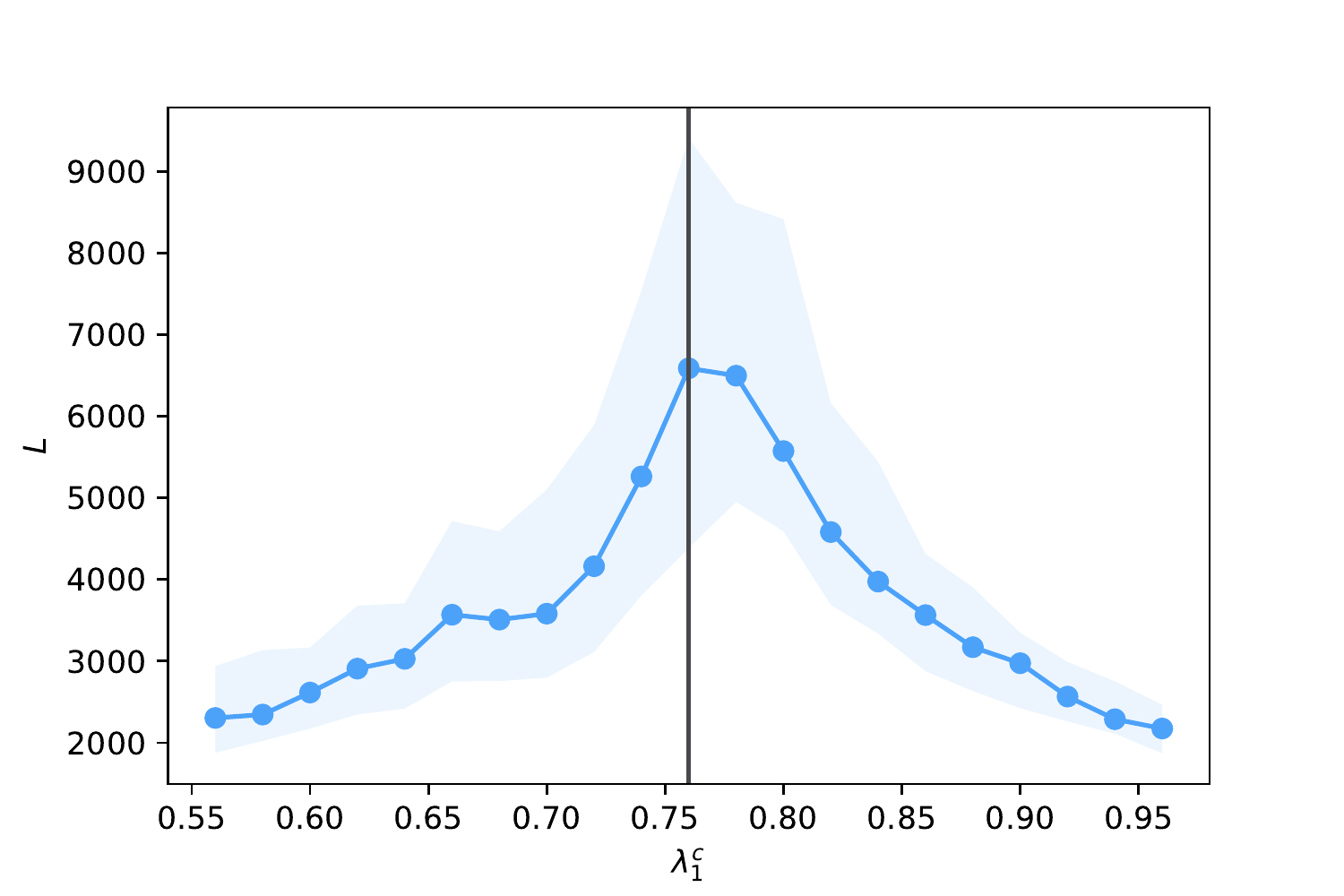}
\caption{We show the lifetime of the SIS process in case of $Q=3$ as function of $\lambda_1$. The vertical line describes the analytical estimation of its critical value. In the simulation we set $\beta_2=\beta_3=0.3$, $\mu_1=\mu_2=\mu_3=0.01$, $N=3 \times 10^{5}$, $m=4$, $\alpha=-2.1$, $\epsilon=10^{-3}$ and run $10^2$ simulations for each data point. We show the $50\%$ confidence intervals in the shaded area and the median with the dots. }
\label{SI_fig:Q3}
\end{figure}

Here we consider the case of three categories. For simplicity let's consider nodes are assigned to the categories at random and that categories have the same size $N^x=N/3$. Also, let's us set the values of $\beta_x$ with $x=[2,3]$, $\mu_x$ with $x=[1,2,3]$, $m=4$, and assume that links are created randomly between categories thus $p=1/3$. In particular, if we set $\beta_2=\beta_3=0.3$, $\mu_1=\mu_2=\mu_3=0.01$, we can use Eq.~\ref{Jacobian} to obtain the critical value of $\lambda_1$. In particular, the general expression of the Jacobian is: 
\begin{widetext}
\be
J=
\begin{bmatrix}
   -\mu_1 & 0 & 0 & p \lambda_1 m & (1-p) \lambda_1 m c_{1,2} & (1-p) \lambda_1 m c_{1,3}\\
      0 & -\mu_2 & 0 & (1-p) \lambda_2 m c_{2,1} & p \lambda_2 m & (1-p) \lambda_2 m c_{2,3} \\
            0 & 0 & -\mu_3 & (1-p) \lambda_3 m c_{3,1} & (1-p) \lambda_3 m c_{3,2} & p \lambda_3 m  \\
          0&  0 & 0 & -\mu_1 + p \beta_1 & (1-p) \beta_1 c_{1,2} & (1-p) \beta_1 c_{1,3} \\
                0 &    0 & 0 & (1-p) \beta_2 c_{2,1} & -\mu_2 + p \beta_2 & (1-p) \beta_2 c_{3,1} \\
                                0 &    0 & 0 & (1-p) \beta_3 c_{3,1} & (1-p) \beta_2 c_{3,2} & -\mu_3 + p \beta_3 \\
\end{bmatrix}
\ee
\end{widetext}
Since the categories have the same size:
\be
c_{x,y}=\frac{N^x}{N-N^y}=\frac{N}{3}\frac{1}{N-\frac{N}{3}}=\frac{1}{2}
\ee
Plugging all the values and solving for $\lambda_1$ we obtain:
\be
\lambda_1^c=\frac{42}{55}
\ee
In Figure~\ref{SI_fig:Q3} we show the comparison between the analytical prediction and the numerical simulations which perfectly matches. 

\end{document}